\documentclass[12pt]{article}
\newfont{\bbd}{msbm10 scaled\magstep1}
\newfont{\frak}{eufm10 scaled\magstep1}
\def\Cbbd{\hbox{\bbd C}}
\def\Lbbd{\hbox{\bbd L}}
\def\Rbbd{\hbox{\bbd R}}
\def\Sbbd{\hbox{\bbd S}}
\def\Zbbd{\hbox{\bbd Z}}
\def\rfrak{\mbox{\frak r}}
\def\Lfrak{\mbox{\frak L}}
\def\Bcal{{\cal B}}
\def\Lcal{{\cal L}}
\def\Mcal{{\cal M}}
\def\Pcal{{\cal P}}
\def\Qcal{{\cal Q}}

\def\e{\hbox{\rm e}}   
\def\i{\hbox{\rm i}}   

\def\phi{\varphi}
\def\eps{\varepsilon}
\def\a{\alpha}
\def\b{\beta}
\def\g{\gamma}

\def\d{\delta}
\def\D{\Delta}
\def\l{\lambda}

\def\om{\omega}

\let\tilde=\widetilde
\def\tr{\mathop{\hbox{\rm tr}}\nolimits}
\def\dd{\partial}
\def\id{\hbox{{1}\kern-.25em\hbox{\rm l}}}
\def\one#1{#1^{\raise5pt\hbox{$\scriptstyle\!\!\!\!1$}}\,{}}
\def\two#1{#1^{\raise5pt\hbox{$\scriptstyle\!\!\!\!2$}}\,{}}
\def\half{\frac{1}{2}}

\def\tbox#1{{\mbox{\tiny\rm #1}}}
\def\bt{B\"acklund transformation}
\def\comment#1{}

\def\Ref#1{(\ref{#1})}
\def\?{(?)\marginpar{|?}}

\def\beq{\begin{equation}}
\def\eeq{\end{equation}}
\def\be{\begin{displaymath}}
\def\ee{\end{displaymath}}
\def\bea{\begin{eqnarray}}
\def\eea{\end{eqnarray}}
\def\bmat{\left(\begin{array}}
\def\emat{\end{array}\right)}
\newcounter{subequation}[equation]
\makeatletter
\expandafter\let\expandafter
\reset@font\csname reset@font\endcsname
\def\subeqnarray{\arraycolsep1pt
    \def\@eqnnum\stepcounter##1{\stepcounter{subequation}%
        {\reset@font\rm(\theequation\alph{subequation})}}
\jot5mm     \eqnarray}

\makeatother
\begin{document}
\begin{center}\LARGE\bf
B\"{a}cklund transformations and Baxter's $Q$-operator
\end{center}
\vskip1cm
\begin{center}
E K Sklyanin
\vskip0.4cm
Steklov Mathematical Institute at St.~Petersburg,\\
Fontanka 27, St.~Petersburg 191011, Russia.\\
E-mail: {\tt sklyanin\symbol{'100}euclid.pdmi.ras.ru}
\end{center}
\vskip2cm
{\bf Abstract.} The course of 5 lectures given at the seminar
``Integrable Systems: from Classical to Quantum'' (Universit\'e de Montr\'eal,
Jul 26 -- Aug 6, 1999) contains a detailed comment on the recently discovered
(Gaudin-Pasquier, 1992) connection between \bt s in the theory of classical
integrable systems on one hand, and Baxter's $Q$-operator for quantum
integrable systems, on the other hand. We restrict our attention to the
systems with finite number of degrees of freedom. Our main illustrative
example is the periodic Toda lattice. We present a general construction of
$Q$-operator for models governed by the $SL(2)$-invariant $R$-matrix and
apply it to our example. We discuss also applications of BT and $Q$-operators
to the separation of variables and theory of special functions.

\vskip3mm
\noindent{}Mathematical Subject Classification 2000: 
{\sf 37J35, 70H06, 81R50} \\
\noindent{}Running title: {\it B\"{a}cklund transformations\ldots}
\newpage
\pagestyle{plain}
\tableofcontents
\section{Introduction}

\subsection{A bit of history}
\label{ss:history}

The B\"{a}cklund and Darboux transformations appeared in XIX century
in the study of the problems of differential geometry. With the advent
of the Inverse Scattering Method in 1960s, their relevance to the 
integrable nonlinear evolution equations was quickly recognized, and the
amount of literature accumulated since then is enormous. See, for example,
the monograph \cite{MS91}. Especially important for these lectures is
the Hamiltonian interpretation of BT discovered by Flaschka and Mclaughlin
\cite{FM76}.

The $Q$-operator belongs to the realm of quantum integrability, and, compared 
to BT, is a relatively new invention.
The $Q$-operator was introduced first by Rodney Baxter in his seminal study 
\cite{Bax72}, see also \cite{Bax82}, 
of the integrable quantum XYZ spin chain as an ingenious device
which allowed to determine the spectrum of the model --- the problem which was
intractable
by other known methods, such as Bethe ansatz. The $Q$-operator is
actually a one-parametric family of operators $Q_\lambda$ commuting with
the Hamiltonians of the integrable system. Its main characteristic property
is that its eigenvalues satisfy certain finite-difference, or, depending
on the integrable model, differential
equation with respect to the parameter $\lambda$, known nowadays as 
{\it Baxter equation}. The Baxter equation, together with appropriate
boundary conditions, provides a one-dimensional multiparameter spectral
problem which allows to determine the spectrum of the commuting
Hamiltonians of the model in question. Thus an originally multidimensional
spectral problem is reduced to a one-dimensional one --- phenonomenon similar
to the separation of variables (SoV). The coincidence is not an acccident. 
Indeed,
as shown in \cite{KS5,KV00}, for classical Hamiltonian systems there exists an
intimate relation between SoV and B\"acklund transformation (BT), 
the latter being
the classical analog of $Q$-operator \cite{KS5,PG92}.

For a long time the XYZ model remained the only model for which a
$Q$-operator
was known. In 1992 Pasquier and Gaudin \cite{PG92} have constructed
a $Q$-operator for the quantum periodic Toda lattice using a somewhat different
approach from Baxter's one. They have described $Q$-operator explicitely,
as an integral operator, and have found an important relation between
the $Q$-operator and the B\"acklund transformation from the classical Toda
chain. Namely, the B\"acklund transformation, as a canonical transformation,
coincides with the classical limit $\hbar\rightarrow0$ of the automorphism
${\cal O}\mapsto Q_\l{\cal O}Q_\l^{-1}$ of the associative algebra of quantum
observables generated by $Q_\l$. The generating function $F_\l(y|x)$ of the
B\"acklund transformation is obtained from the semiclssical asymptotics
$ Q_\l(y|x)\sim \exp(i\hbar^{-1}F_\l(y|x))$ of the kernel $Q_\l(y|x)$
of $Q_\l$ considered as an integral operator in the coordinate representation.

Later on, in \cite{BLZ} Bazhanov, Lukyanov and Zamolodchikov
gave a boost to the original Baxter's idea of constructing $Q_\l$ as the trace
of the monodromy matrix constructed of Lax operators corresponding to a special
representation of the relevant quantum group in the auxiliary space.
Setting the problem in the context of representation theory for quantum groups,
they have taken as such representation of $\widehat{sl}_q(2)$ the so-called
$q$-oscillator representation and have managed to construct a pair of
$Q$-operators for
the massless sine-Gordon quantum field theory in a periodic box.
It seems that the same $q$-oscillator representation allows to construct
a $Q$-operator for any integrable model governed by the quantum group
$\widehat{sl}_q(2)$, see \cite{AF97}.

In the paper \cite{KSS98} devoted to construction of a
$Q$-operator for the so-called DST (dimer self-trapping) model (a degenerate
case of XXX magnetic chain) the combination of the approaches due to
Baxter ($Q_\l$ as a trace of monodromy)
and Pasquier-Gaudin ($Q_\l$ as an integral operator) allowed to describe
the structure of the $Q$-operator in the greatest detail. Besides an
explicit expression for the kernel of the integral operator $Q_\l$ in
several equivalent forms one can calculate explicitely the matrix elements
of $Q_\l$ in the natural monomial basis.

In the last years a considerable progress is achieved in the understanding
of the Hamiltonian properties of the BT for the classical integrable systems
which parallel those of the $Q$-operator in the quantum case.
It is worth noticing that the classical counterparts of some of the 
properties of the quantum $Q$-operator were unknown before. As an example one
can mention the so-called {\it spectrality property} of BT discovered in
\cite{KS5} which corresponds to Baxter's finite-different equation for
the quantum case. Baxter's construction of $Q_\l$ as a trace of monodromy
matrix has led to a new construction of BT from symplectic leaves of
the quadratic $r$-matrix Poisson bracket \cite{Skl51}.

A special topic actively studied in the last years
is the relation of BT and SoV  which turns out to be twofold.
On one hand SoV can be obtaind from a composition of BT or $Q$, see
\cite{KS5}.
On the other hand, a BT can, in turn, be obtained as the transformation
intertwining a pair of SoV \cite{KV00}. In this case, the quantum 
interpretation is not found yet.

The growing interest in studying various properties of $Q$-operator for a 
variety of quantum integrable systems
is indicated by a surge of recent publications, see \cite{Der,Pro,Sm-rt}. 

\subsection{Plan of lectures}
\label{ss:plan}

In these lectures I will concentrate on the parallels between the 
\bt\ for the classical Hamiltonian sytems one one hand and the $Q$-operator
for the quantum integrable systems, on the other hand.
As it frequently happens, when two theories merge after having been 
developed for a considerable time independently, the resulting 
cross-fertilization is quite useful for both. The most recent example is
the classical $r$-matrix and Lie-Poisson groups \cite{STS83,Sem} whose
invention was inspired by the quantum theory.

We shall restrict our attention to the systems with a finite number of degrees
of freedom (pure quantum mechanics, no field theory) and 
put special stress on Hamiltonian mechanics which is essential for
quantization. All the new notions and techniques  will be 
introduced on the example of the periodic Toda lattice and accompanied
with a short discussion of possible generalizations.

The lectures can be considered as an extended commentary to the paper
by Pasquier and Gaudin \cite{PG92} accompanied by the original results
obtained by V.~Kuznetsov and myself \cite{KS5,KSS98,Skl51,Skl52}.  

\section{Classical periodic Toda lattice}

\subsection{Description of the model}
\label{ss:model}

The periodic Toda lattice \cite{Toda,Moe76} is a system of $n$ degrees of
freedom described in terms of canonical coordinates $x\equiv(x_1,\ldots,x_n)$
and momenta $X\equiv(X_1,\ldots,X_n)$ having the standard Poisson brackets
\beq
  \{X_j,X_k\}=\{x_j,x_k\}=0, \qquad \{X_j,x_k\}=\d_{jk}.
\label{eq:Pb}
\eeq

In what follows we always denote canonical coordinates with small letters,
e.g.\ $x$, $y$, $s$, $t$, $\phi$ and the corresponding canonical momenta with
the respective capital letters: $X$, $Y$, $S$, $T$, $\Phi$. Such a convention
helps to deal with several sets of canonical variables.

The physical Hamiltonian $H$ of the Toda lattice
\beq
 H=\sum_{j=1}^n \left(\half X_j^2+\e^{x_{j+1}-x_j}\right)
\label{eq:def-H}
\eeq
describes the system of $n$ one-dimensional non-relativistic particles of
equal mass interacting  via exponential potential between the nearest
neighbors. In formulas like \Ref{eq:def-H} we always assume the periodicity
convention: $j+n\equiv j$. The Hamiltonian $H$ is thus invariant with respect
to the translation $j\mapsto j+1$, hence the name `periodic Toda lattice'.

The equations of motion $\dot f=\{H,f\}$ corresponding to the Hamiltonian
\Ref{eq:def-H} are
\begin{subeqnarray}
  \dot x_j&=&X_j, \\
  \dot X_j&=&-\e^{x_j-x_{j-1}}+\e^{x_{j+1}-x_j}.
\label{eq:eqs-mot}
\end{subeqnarray}

\subsection{Integrability}
\label{ss:integrability}

The periodic Toda lattice is an example of a {\it completely integrable
Hamiltonian system} in the Liouville-Arnold sense \cite{Arnold}.
It means that the Hamiltonian $H$
\Ref{eq:def-H} is an element of a ring generated by $n$ independent
Hamiltonians $H_1$,\ldots,$H_n$ which commute
\beq
   \{H_{j_1},H_{j_2}\}=0
\label{eq:comH}
\eeq
with respect to the Poisson
bracket \Ref{eq:Pb}. As a consequence, the Hamiltonian flow $\dot f=\{H,f\}$
leaves the Hamiltonians $H_j$ invariant $\dot H_j=0$ and, therefore,
leaves invariant the level manifolds $P_h$
obtained by fixing the values of the Hamiltonians $H_j=h_j$.

The fundamental result in the theory of Hamiltonian integrable systems is
the Liouville theorem \cite{Arnold} which claims that the level manifolds
$P_h$, if compact, are diffeomorphic to $n$-dimensional tori $T^n$.
Moreover, there exist such canonical {\it action-angle} variables $\Phi$,
$\phi$ that the action variables $\Phi_j$ are functions of the Hamiltonians
$H_1$,\ldots,$H_n$, and the Hamiltonian flows linearize in the angle
variables: $\{H_j,\dot\phi_k\}=\om_{jk}(\Phi)$.

The easiest way to demonstrate the integrability of the periodic Toda lattice
is to make use of the Inverse Scattering method (or, the Isopectral
Deformation method) in its Hamiltonian version,
see for example \cite{FT87} and Prof.\ J.~Harnad's lectures in this volume.
Within the ISM framework, the commuting
Hamiltonians are obtained from the spectral invariants of the
{\it Lax matrix} $L(u; X,x)$ which is a square matrix of order $N$
(generally speaking, different from the number of degrees of freedom $n$)
depending on a complex parameter $u$ called {\it spectral parameter} and
whose matrix elements are functions on the phase space.

The spectral invariants $t_k(u)$ of $L(u)$ defined as the coefficients
of the {\it characteristic polynomial}
\beq
  W(v,u)\equiv \det\bigl(v-L(u)\bigr)=v^N+\sum_{k=1}^N (-1)^k v^{N-k}t_k(u)
\label{eq:def-W}
\eeq
are elementary symmetric polynomials of
the eigenvalues of $L(u)$, or, in terms of matrix elements of $L(u)$,
sums of principal minors of order $k$ (determinants of submatrices of $L(u)$
of order $k$ whose diagonal is contained in the diagonal of $L(u)$).
For example, $t_1(u)=\tr L(u)$, $t_N(u)=(-1)^N\det L(u)$. The commuting
Hamiltonians $H_j$ are usually obtained as coefficients of $t_k(u)$ when
$t_k(u)$ are polynomials in $u$, or coefficients of expansions of $t_k(u)$
in some other bases of functions of $u$ (e.g.\ trigonometric or elliptic
ones).

Proving the commutativity \Ref{eq:comH} of the Hamiltonians $H_j$ is
equivalent thus to proving the commutativity
\beq
 \{t_{k_1}(u_1),t_{k_2}(u_2)\}=0
\label{eq:comm-t}
\eeq
of the spectral invariants $t_k(u)$.
The following theorem due to Babelon and Viallet  provides a
technical mean to do it. As proven in \cite{BV90}, the
the commutativity \Ref{eq:comm-t} of the spectral invariants of $L(u)$
is equivalent to the
existence of a so-called $\rfrak$-matrix represenation for the Poisson brackets
$\{L_{a_1b_1}(u_1),L_{a_2b_2}(u_2)\}$ between the matrix elements
$L_{ab}(u)$. To write down the representation in a compact form,
we introduce the tensor product notation
\beq
  \one L\equiv L\otimes\id, \qquad
  \two L\equiv\id\otimes L,
\eeq
where $\id$ is the unit matrix of order $N$. Respectively,
$\{\one L(u_1),\two L(u_2)\}$ is the matrix of order $N^2\times N^2$ of all
Poisson brackets between the matrix elements of $L(u_1)$ an $L(u_2)$.
The theorem of Babelon and Viallet claims that the commutativity
\Ref{eq:comm-t} is equivalent to the existence of two matrices,
$\rfrak_{12}$ and $\rfrak_{21}$, of order $N^2\times N^2$
such that the equality
\beq
  \{\one L(u_1),\two L(u_2)\}
   = [\rfrak_{12},\one L(u_1)]
    -[\rfrak_{21},\two L(u_2)]
\label{eq:BV}
\eeq
holds for any $u_1$, $u_2$. Note, that, $\rfrak_{12}$ and $\rfrak_{21}$
depend on $u_1$ and $u_2$, and, generally speaking, contain the dynamical
variables $X$, $x$.
Actually, one can always choose the
$\rfrak$-matrices in such a way that
$\rfrak_{21}(u_1,u_2)=\Pcal_{12}\rfrak_{12}(u_2,u_1)\Pcal_{12}$
where $\Pcal_{12}$ is
the permutation matrix: $\Pcal_{12}x\otimes y=y\otimes x$.

Speaking again about
the periodic Toda chain, in order to construct the commutative
Hamiltonians, we have to produce a Lax matrix and the corresponding
$\rfrak$-matrices. There are at least two possible Lax matrices for the periodic Toda
chain, one of order $2\times2$, another one of order $n\times n$,
see \cite{FT87,Toda,Moe76,AHH90}.
In this subsection we shall work with the $2\times2$ matrix.

The $2\times2$ Lax matrix (or, monodromy matrix \cite{FT87})
$L(u;X,x)$ is defined as the product of local Lax matrices
$\ell_j(u)$ depending on the variables $X_j$ and $x_j$ only:
\beq
 L(u)=\ell_n(u)\ldots\ell_2(u)\ell_1(u),
\label{eq:toda-Lll}
\eeq
\beq
 \ell_j(u)\equiv\ell_j(u;X_j,x_j)=\left(\begin{array}{cc}
     u+X_j & -\e^{x_j} \\
     \e^{-x_j} & 0 \end{array}\right).
\label{eq:def-ell}
\eeq

The characteristic polynomial of $L(u)$ is quadratic in $v$ having thus two
spectral invariants: $t_1(u)$ and $t_2(u)$. However,
$t_2(u)=\det L(u)\equiv1$ by virtue of $\det\ell(u)=1$ which leaves
$t(u)\equiv t_1(u)=\tr L(u)$ as the only nontrivial spectral invariant:
\beq
  W(u,v)\equiv\det\bigl(v-L(u)\bigr)=v^2-t(u)v+1.
\eeq

The Hamiltonians $H_j$ are obtained then from the expansion of $t(u)$:
\beq
 t(u)=u^n+H_1u^{n-1}+\ldots+H_n.
\label{eq:def-Hj}
\eeq

In particular, $H_1=X_1+\ldots+X_n$ is the total momentum.
It is easy to see that the physical Hamiltonian
\Ref{eq:def-H} is given by the formula $H=\half H_1^2-H_2$ and thus
belongs to the polynomial ring generated by $H_1$,\ldots,$H_n$.

To prove the commutativity \Ref{eq:comH} of the Hamiltonians $H_j$, or,
equivalently, the commutativity
\beq
  \{t(u_1),t(u_2)\}=0
\eeq
of their generating function $t(u)$, it is sufficient to find the
corresponding $\rfrak$-matrices. Actually, what we are able to prove
is a much more special representation \cite{FT87} for the left-hand-side
of \Ref{eq:BV}:
\beq
    \{\one L(u),\two L(v)\}=[r_{12}(u-v),\one L(u)\two L(v)]
\label{eq:Pb-LL}
\eeq
where
\beq
r_{12}(u_1-u_2)=\frac{\Pcal_{12}}{u_1-u_2}
\label{eq:r-toda}
\eeq
is the $SL(2)$-invariant solution to the classical Yang-Baxter equation
\beq
 [r_{12}(u),r_{13}(u+v)]+[r_{12}(u),r_{23}(v)]+[r_{13}(u+v),r_{23}(v)]=0.
\eeq

One can easily transform the formula \Ref{eq:Pb-LL} to the Babelon-Viallet
form by setting in \Ref{eq:BV}
\begin{subeqnarray}
\rfrak_{12}&=&
   \half\bigl(r_{12}(u_1-u_2)\two L(u_2)+\two L(u_2)r_{12}(u_1-u_2)\bigr), \\
\rfrak_{21}&=&
   -\half\bigl(r_{12}(u_1-u_2)\one L(u_1)+\one L(u_1)r_{12}(u_1-u_2)\bigr).
\end{subeqnarray}

To prove \Ref{eq:Pb-LL} we shall make use of the so-called
{\it comultiplication} property of the quadratic Poisson bracket.
It is a simple exercise to verify that if two Lax matrices $L_1(u)$ and
$L_2(u)$ defined, respectively, on different phase spaces $P_1$ and $P_2$
satisfy each the identity \Ref{eq:Pb-LL} then their matrix product
$L(u)=L_1(u)L_2(u)$ defined on the direct product $P_1\times P_2$
satisfies the same identity. The proof uses nothing but the identity
\Ref{eq:Pb-LL} for $L_1$ and $L_2$, and the identity
\beq
  \{\one L_1(u_1),\two L_2(u_2)\}=0.
\eeq

It is sufficient thus to verify the identity \Ref{eq:Pb-LL} for the local
Lax matrices $\ell_j(u)$ given by \Ref{eq:def-ell} which is a matter of
a direct calculation.

Strictly speaking, to establish the integrability, besides the proof
prove of the commutativity of Hamiltonians $H_j$ we need to prove their
independence. For a proof, see \cite{RS-viniti}. It is also possible to
verify that, modulo center-of-mass motion, which is easily separated,
the level manifolds $P_h$ are compact and thus, by virtue of Liouville
theorem are isomorphic to tori.

\subsection{Quadratic Poisson bracket}
\label{ss:qpb}

Before starting the discussion of \bt{s} we need to learn some more facts
about the $r$-matrix quadratic Poisson bracket \Ref{eq:Pb-LL} and the
class of integrable models it generates.

Let us suppose that $2\times2$ matrix
$L(u)$ is a polynomial in $u$ of degree $n$
\beq
  L(u;X,x)=L^{(n)}u^n+L^{(n-1)}u^{n-1}+\ldots+L^{(0)}
\label{eq:poly-L}
\eeq
and regard the equality \Ref{eq:Pb-LL}
with the $r$-matrix given by \Ref{eq:r-toda}
as introducing a Poisson bracket on
the $4(n+1)$ variables $L^{(j)}_{ab}$, $j=0,\ldots,n$, $a,b=1,2$.
For the sake of simplicity we think of $L^{(j)}_{ab}$ as complex variables
and do not consider here the question of choosing an appropriate
$\ast$-conjugation.

It is easy to see, that, despite the denominator $(u_1-u_2)$ present in the
$r$-matrix \Ref{eq:r-toda}, the right-hand-side of \Ref{eq:Pb-LL} is
polynomial both in $u_1$ and $u_2$ because of the identity
$[\Pcal,L\otimes L]=0$ which nullifies the numerator for $u_1=u_2$.
According to a theorem by Sophus Lie \cite{Wei}, for any Poisson bracket
there exist local coordinates
$(X,x,c)$ such that $X$ and $x$ are canonical \Ref{eq:Pb},
and $c$ are central, that is
\be
\{c_j,c_k\}=\{c_j,X_k\}=\{c_j,x_k\}=0.
\ee

To obtain a symplectic manifold which can serve as a phase space for a
mechanical system, one needs thus to restrict the Poisson bracket onto
a level manifold of its central (or Casimir) functions.

In case of the bracket \Ref{eq:Pb-LL} the Casimir functions can be
found easily. First, the leading coefficient $L^{(n)}$ provides 4 casimirs.
More casimirs are given by the coefficients of the determinant $\det L(u)$.
Being, generally speaking, a polynomial of degree $2n$, the determinant
has $(2n+1)$ coefficient but its leading coefficient coincides with
$\det L^{(n)}$ which gives us only $2n$ new casimirs. In total, we have
$(2n+4)$ casimirs which corresponds to the level manifolds of dimension
$4(n+1)-(2n+4)=2n$. To show that there are no more casimirs, it is sufficient
to construct an example of $2n$-dimensional symplectic leaf of the bracket
\Ref{eq:Pb-LL}.

The tool for constructing such examples is the {\it comultiplication property}
of the bracket \Ref{eq:Pb-LL} mentioned in the subsection
\ref{ss:integrability}. It allows to build multidimensional symplectic
leaves from simpler blocks. The simplest, 0-dimensional simplectic leaf
of the bracket \Ref{eq:Pb-LL} is given by a constant matrix $L(u)\equiv K$.
The next most natural choice is to take a linear polynomial in $u$ with the
unit matrix as the leading coefficient:
\beq
   \ell^{\tbox{XXX}}(u)=u\id+\Sbbd, \qquad
   \Sbbd=\bmat{cc} S_3 & S_1-\i S_2 \\ S_1+\i S_2 & -S_3 \emat.
\label{eq:lXXX}
\eeq

Substituting \Ref{eq:lXXX} for $L(u)$ into \Ref{eq:Pb-LL} we
obtain for $S_\a$ the Poissonian algebra isomorphic to the Lie algebra
$sl_2$:
\beq
   \{S_\a,S_\b\}=-\i\sum_{\g=1}^3 \eps_{\a\b\g}S_\g,
\label{eq:sl2}
\eeq
$\eps_{\a\b\g}$ being the standard antisymmetric tensor.
The Poisson bracket \Ref{eq:sl2} has the Casimir function
$C=S_1^2+S_2^2+S_3^2$, and its generic symplectic leaves
$C=\hbox{\rm const}\neq0$ are 2-dimensional spheres.

Noting that, due to the fact that the $r$-matrix \Ref{eq:r-toda} depends
only on the difference $(u_1-u_2)$, the shift of the spectral parameter
$u\mapsto u-c$ is an automorphism of the poissonian algebra \Ref{eq:Pb-LL}.
Therefore, taking a direct product of
$n$ copies of the triplets $S_\a$ restricted to the
level surfaces $C=\rho^2_j$, $j=1,\ldots,n$ we obtain a $2n$-dimensional
symplectic leaf of the bracket \Ref{eq:Pb-LL} given by the product
\beq
   L^{\tbox{XXX}}(u)=
K\ell_n^{\tbox{XXX}}(u-c_n)\ldots\ell_1^{\tbox{XXX}}(u-c_1).
\label{eq:L-XXX}
\eeq

Note that the number of parameters contained in the symplectic leaf is
is $(2n+4)$ where 4 comes from the constant matrix $K$ and rest from
$n$ casimirs $\rho_j$ and $n$ shifts $c_j$, $j=1,\ldots,n$. The parameters
are easily identified with
$L^{(n)}=K$ and the zeroes of the determinant
$\det L(u)=\det K\prod_j (u-c_j-\rho_j)(u-c_j+\rho_j)$.
We are thus led to the conclusion that the constructed symplectic leaf
is in fact the generic leaf for the bracket \Ref{eq:Pb-LL}.

The Lax matrix \Ref{eq:L-XXX} defines an integrable system known as
the {\it inhomogeneous Heisenberg magnetic chain} \cite{FT87,QNCL84}.
All other integrable models associated with the Poisson bracket \Ref{eq:Pb-LL}
and the $sl_2$-invariant $r$-matrix \Ref{eq:r-toda} can be obtained
from degenerations of the Lax matrix \Ref{eq:L-XXX}.

To describe some important degenerations of \Ref{eq:L-XXX} let us parametrize
the spin components $S_\a$ in \Ref{eq:lXXX} using a pair of
canonical variables $(X,x)$:
\beq
   \ell^{\tbox{XXX}}(u)=
   \bmat{cc} u+xX-\rho & -x^2X+2\rho x \\ X & u-xX+\rho \emat
\label{eq:lxxx-Xx}
\eeq

Multiplying $\ell^{\tbox{XXX}}(u)$ from the right
by the diagonal matrix $\mbox{\rm diag}(1,-1/(2\rho))$
and performing the shift $u\mapsto u+\rho$ (note that these
are legal operations which do not change the Poisson bracket \Ref{eq:Pb-LL})
we are capable to take the limit $\rho\rightarrow\infty$.
The result is the Lax matrix
for the so-called dimer-self-trapping (DST) model \cite{KSS98}
\beq
   \ell^{\tbox{DST}}(u)=
   \bmat{cc} u+xX & -x \\ X & -1 \emat.
\label{eq:lDST}
\eeq

Note that the determinant $\det\ell^{\tbox{DST}}(u)=-u$ is linear
in $u$. A further degeneration of DST model produces the Toda lattice.
To this end, one multiplies $\ell^{\tbox{DST}}(u)$
from the right by the matrix $\mbox{\rm diag}(1,a^{-1})$ and, after making
substitutions
$u\mapsto u-a$, $x\mapsto a\e^x$, $X\mapsto \e^{-x}(1+a^{-1}X)$,
obtains in the limit $a\rightarrow\infty$ the unimodular Lax matrix
\Ref{eq:def-ell} for the Toda lattice.

More symplectic leaves can be obtained by applying the automorphism
$\ell(u)\mapsto\check{\ell}(u)\equiv\ell^{-1}(-u)$ of the $r$-matrix Poisson
algebra \Ref{eq:Pb-LL} to $\ell^{\tbox{DST}}(u)$ and
$\ell^{\tbox{Toda}}(u)$. Up to a scalar factor we have:
\beq
  \check{\ell}^{\tbox{DST}}(u)\sim
   \bmat{cc} 1 & -x \\ X & u-xX \emat, \qquad
  \check{\ell}^{\tbox{Toda}}(u)\sim
   \bmat{cc} 0 & -\e^{x} \\ \e^{-x} & u-X \emat
\label{eq:check-l}
\eeq
(on ${\ell}^{\tbox{XXX}}(u)$ the automorphism acts trivially:
$\check{\ell}^{\tbox{XXX}}(u)\sim{\ell}^{\tbox{XXX}}(u)$).

{\bf Conjecture.} {\it Any symplectic leaf $L(u)$ of the
quadratic $r$-matrix Poisson bracket \Ref{eq:Pb-LL}
which is polynomial in $u$ can be decomposed
(in a non-unique way, of course) into a product of a constant matrix $K$ and
linear matrix polynomials of the form $\ell^{\tbox{XXX}}(u-c)$,
$\ell^{\tbox{DST}}(u-c)$,
$\check\ell^{\tbox{DST}}(u-c)$, $\ell^{\tbox{Toda}}(u-c)$,
$\check\ell^{\tbox{Toda}}(u-c)$.
}

In case of a generic symplectic leaf the factorization \Ref{eq:L-XXX}
in terms of $\ell^{\tbox{XXX}}(u-c)$ only should suffice.
The difficult part
is to analyze the degenerate cases when the leading coefficient
$L^{(n)}$ is a degenerate matrix and/or the degree of $\det L(u)$ is less
than $2n$. Hopefully, one of the readers will provide a proof soon.

One can find more information about the properties of the $r$-matrix Poisson
bracket \Ref{eq:Pb-LL} in the papers \cite{FT87,Sem,STS83} as well as in the
lectures by Harnad and Reshetikhin in the present volume. 

\subsection{\bt\ and its properties}
\label{ss:bt}

In this section we start to study the \bt\ for the periodic Toda lattice.
The \bt\ $\Bcal_\l$ depending on a complex parameter $\l$
is defined as the mapping
from the variables $(X,x)$ to $(Y,y)$
given implicitely by the equations
\begin{subeqnarray}
 X_j&=&\e^{x_j-y_j}+\e^{y_{j+1}-x_j}-\l, \\
 Y_j&=&\e^{x_j-y_j}+\e^{y_j-x_{j-1}}-\l.
\label{eq:XY-toda}
\end{subeqnarray}

The equations \Ref{eq:XY-toda} are algebraic in momenta and
exponents of coordinates. Resolving (\ref{eq:XY-toda}a) with respect to
$\e^{y_{j+1}}$:
\beq
   \e^{y_{j+1}}= \e^{x_j}(X_j+\l)-\e^{2x_j-y_j}
\label{eq:iter-y}
\eeq
and iterating the equation \Ref{eq:iter-y} for $j=1,2,\ldots,n$ we finally
arrive to a {\it quadratic} equation for $\e^{y_1}$ which implies that
the transformation $\Bcal_\l$ is a {\it two-valued} algebraic function
in terms of $X$, $\e^x$.
Fortunately, for all our purposes the simple
implicit formulas \Ref{eq:XY-toda} are sufficient.

Another attractive feature of the equations \Ref{eq:XY-toda}
is their {\it locality}: they involve only the variables with the indices
differing by $0$ and $1$. Note that even for real $\l$ resolving the
equations \Ref{eq:XY-toda} can produce complex values of $Y$ and $y$.
To avoid complications, we shall not make attempt to study the reality
conditions and treat both $(X,x)$ and $(Y,y)$ as complex variables, in
the spirit of {\it algebraic integrability} \cite{AvM89}.

We start the list of properties of the \bt\ with noting its
{\it canonicity}: the variables $(Y_j,y_j)$ are canonical.
It can be seen from the fact
that the equations \Ref{eq:XY-toda} can be written down in the form
\beq
  X_j=\frac{\dd F_\l}{\dd x_j}, \qquad
  Y_j=-\frac{\dd F_\l}{\dd y_j}.
\label{eq:XYxy}
\eeq
where $F_\l(y;x)$
\beq
 F_\l(y;x)=\sum_{i=1}^n \bigl(\e^{x_j-y_j}-\e^{y_{i+1}-x_j}-\l(x_j-y_j)\bigr)
\label{eq:F-toda}
\eeq
is the generating function \cite{Arnold} of the canonical transformation.

The next property is the {\it invariance of Hamiltonians}:
\beq
    H_j(X,x)=H_j(Y,y), \qquad j=1,\ldots,n.
\label{eq:inv-H-cl}
\eeq

Though the invariance of physical Hamiltonian $H$ \Ref{eq:def-H}
can be proved by a direct
calculation \cite{Toda,Gau83}, to prove the invariance of the whole set of
commuting Hamiltonians $H_j$ we will need some more effective technique.
The easiest way is to make use of
Inverse Scattering Method explained in section \ref{ss:integrability}.
The invariance of $H_j$ under $\Bcal_\l$ is
equivalent then
to the invariance of the spectrum of $L(u)$ which implies that there
exists an invertible matrix $M(u,\l)$ such that
\beq
   M(u,\l)L(u;Y,y)=L(u;X,x)M(u,\l),\qquad \forall u\in\Cbbd.
\label{eq:ML}
\eeq

Such a matrix is called {\it Darboux matrix}, and the tranformation of
$L$ given by \Ref{eq:ML} is called {\it Darboux transformation}
\cite{MS91}.

In our case, due to the factorization \Ref{eq:toda-Lll} of $L(u)$ into
local Lax matrices $\ell_j(u)$, we can be more specific about the structure
of Darboux transformation. Setting $M_1(u,\l)\equiv M(u,\l)$ we introduce
matrices $M_{j+1}(u,\l)$, $j=1,\ldots,n-1$ inductively as
\be
  M_{j+1}(u,\l)=\ell_j(u;X_j,x_j)M_j(u,\l)\ell_j^{-1}(u;Y_j,y_j)
\ee
(note that for $j=n$, due to the periodicity $n+1\equiv1$
we recover the equality \Ref{eq:ML}). The global transformation \Ref{eq:ML}
takes thus form of the local {\it gauge transformation}
\beq
 M_{j+1}(u,\l)\ell_j(u;Y_j,y_j)=\ell_j(u;X_j,x_j)M_j(u,\l).
\label{eq:gauge-toda}
\eeq

The converse is also true: from \Ref{eq:gauge-toda} it follows that the
spectrum of $L(u)$ is preserved. To prove the invariance of Hamiltonians
\Ref{eq:inv-H-cl} it is sufficient thus to find the matrices $M_j$
satisfying \Ref{eq:gauge-toda}.
Using the equations \Ref{eq:XY-toda} it is easy to verify that
\Ref{eq:gauge-toda} is satisfied with the following matrices
\cite{Gau83}:
\beq
 M_j(u,\l)
 =\left(\begin{array}{cc}
   u-\l+\e^{y_j-x_{j-1}} & -\e^{y_j} \\
   \e^{-x_{j-1}} & -1
 \end{array}\right).
\label{eq:def-M-toda}
\eeq

The two properties: canonicity and invariance of Hamiltonians constitute
the definition of what is called an {\it integrable map} \cite{Ves91}.
It can be considered as a discrete-time analog of integrable hamiltonian flow.
Veselov \cite{Ves91} has proved a discrete-time analog of Liouville
theorem which claims that in the action-angle variables $(\Phi,\phi)$
any integrable map
acts as a shift $\phi_j\mapsto\phi_j+\Omega_j(\Phi)$,
or, speaking more precisely, as a collection of shifts due to the
multivaluedness of algebraic mappings. Applying the theorem to the case of
\bt\ depending on a parameter $\l$ and noting that shifts on the Liouville
torus commute we obtain as an immediate consequence the {\it commutativity}
of BT
\beq
   \Bcal_{\l_1}\circ\Bcal_{\l_2}=\Bcal_{\l_2}\circ\Bcal_{\l_1}.
\label{eq:commBT}
\eeq

Note that a direct proof of commutativity of BT is not a simple task,
see for example \cite{Toda,HS}. It is trivialized in our case entirely due
to the fact that from the very beginning we are working in the hamiltonian
context.

The last property in our list is {\it spectrality}. It was discovered rather
recently \cite{KS5}, and the main motivation for its existence comes from
the quantum case, see section \ref{ss:Q-prop}.

Let us introduce the quantity $\mu$ which is, in a sense, canonically
conjugated to $\l$:
\beq
  \mu\equiv \frac{\dd F_\l}{\dd\l}=\sum_{j=1}^n(x_j-y_j).
\label{eq:def-mu-toda}
\eeq

The spectrality of BT means that the pair $(\e^\mu,\l)$ lies on the spectral
curve of the Lax matrix. Since $\det L(u)=1$ it means that both $\e^\mu$
and $\e^{-\mu}$ are eigenvalues of $L(\l)$
\beq
  W(\e^{\pm\mu},\l)\equiv\det\bigl(\e^{\pm\mu}-L(\l)\bigr)=0
\label{eq:spectr}
\eeq
(it does not matter if we take $L(\l;X,x)$ or $L(\l;Y,y)$ since they are
isospectral).

The property of spectrality of BT still remains somewhat mysterious and
certainly needs more research to uncover its algebraic and geometric meaning.
The main drawback of the present definition
is its being formulated in quite noninvariant terms of generating function
$F_\l$.

To prove \Ref{eq:spectr}
it suffices to show that, say $\e^\mu$ is an eigenvalue of the matrix
$L(\l;Y,y)$. We shall construct explicitely the corresponding eigenvector
$\om_1$:
\beq
 L(\l;Y,y)\om_1=\e^{\mu}\om_1.
 \label{eq:Lom-toda}
\eeq

From \Ref{eq:def-M-toda} it follows that
$\det\bigl(M_j(u,\l)\bigr)=\l-u$. It is easy to see that
for $u=\l$ the matrix $M_j(\l,\l)$ degenerates into a projector
\beq
  M_j(\l,\l)=\bmat{c} \e^{y_j} \\ 1
         \emat
         \begin{array}{cc}
              (\e^{-x_{j-1}} &  -1) \\
              {} & {}
         \end{array}
\eeq
and, as a consequence, has the unique, up to a scalar
factor, null-vector
\beq
\om_j=\left(\begin{array}{c}\e^{x_{j-1}} \\ 1 \end{array}\right), \qquad
  M_j(\l,\l)\om_j=0.
\eeq

Using the identity \Ref{eq:ML} with $M\equiv M_1$
we conclude that
\beq
M_1(\l,\l)L(\l;Y,y)\om_1=0
\eeq
which, combined with the uniqueness of the null-vector $\om_1$ of
$M_1$, implies that $\om_1$ is an eigenvector of $L(\l;Y,y)$. To
determine the corresponding eigenvalue, we apply the same argument
to the identity \Ref{eq:gauge-toda} obtaining the equality
$M_{j+1}(\l,\l)\ell_j(\l;Y_j,y_j)\linebreak[0]\om_j=0$
from which it follows that
$\ell_j(\l;Y_j,y_j)\om_j\sim\om_{j+1}$. The direct calculation shows
that
\beq
 \ell_j(\l;Y_j,y_j)\om_j=\e^{x_{j-1}-y_j}\om_{j+1}.
\label{eq:vacuum-ell}
\eeq
It remains only to use the formulae \Ref{eq:toda-Lll} and
\Ref{eq:def-mu-toda} to arrive finally at \Ref{eq:Lom-toda}.

An alternative variant of the proof, more close to what we shall use
in the quantum case (see section \ref{ss:Baxter-eq}) is to introduce
a gauge transformation with a triangular matrix $N_j$:
\beq
  \widehat\ell_j\equiv N_{j+1}^{-1}\ell_j(\l;Y_j,y_j)N_j
  =\bmat{cc} \e^{y_j-x_{j-1}} & 0 \\
              \e^{-y_j} & \e^{x_{j-1}-y_j}
    \emat, \qquad
   N_j=\bmat{cc} 1 & \e^{x_{j-1}} \\ 0 & 1 \emat
\label{eq:triangN}
\eeq
(note that $\om_j$ coincides with the second column of $N_j$).

The result, as expected, is
\beq
  t(\l)=\tr\ell_n(\l)\ldots\ell_1(\l)
       =\tr\widehat\ell_n\ldots\widehat\ell_1
       =\e^\mu+\e^{-\mu}.
\label{eq:spectrality}
\eeq

We conclude this section with a remark on using BT for generating solitons
which is the main application of BT to the integrable nonlinear evolution
equations \cite{FM76,HS,Toda}. We are following here the argument by Gaudin
\cite{Gau83}.
Let us apply the \bt\ \Ref{eq:XY-toda} to the vacuum state $X_j=x_j=0$.
The equations \Ref{eq:XY-toda} turn into
\begin{subeqnarray}
  0&=&\e^{-y_j}+\e^{y_{j+1}}-\l, \\
  Y_j&=&\e^{-y_j}+\e^{y_j}-\l.
\end{subeqnarray}

Concentrating on the first equation (the second equation describes the
time evolution $Y_j=dy_j/dt$ with respect to the hamiltonian $H$)
we introduce the parametrization:
$\l=2\cosh\kappa$, $\e^{y_0}=\cosh(\a+\kappa)/\cosh\a$.
The general solution can be now written as
\beq
    \e^{y_j}=\frac{\cosh(\a+\kappa(j+1))}{\cosh(\a+\kappa j)}.
\label{eq:soliton}
\eeq

In case of the infinite lattice,
when $j\in\Zbbd$, the formula \Ref{eq:soliton} describes a soliton solution.
Note, however, that the solution \Ref{eq:soliton} has different asymptotics
$\e^{y_j}\rightarrow\e^{\pm\kappa}$
as $j\rightarrow\pm\infty$, satisfying thus the boundary conditions different
from those for the vacuum state. As a result, the energy and values
of other integrals of motion for the soliton solution differ from those
for the vacuum.

The situation is quite different in the periodic case.
The periodicity condition $y_{n+1}=y_1$ can be satisfyed in two ways.
The first one leads to the quantization of the parameter:
$\kappa n\in\pi\i \Zbbd$ and is inacceptable if we want to keep $\l$ free.
Besides, in this way we get a complex solution for $\e^{y_j}$.
Another option is to fix the free parameter $\a$ by setting $\a=\pm\infty$,
which gives us another vacuum state $\e^{y_j}=\e^{\pm\kappa}$ having
the same values of Hamiltonians as the vacuum.

The fact that BT in the periodic case does not produce solitons and
always preserves the integrals of motion may dissapoint those accustomed
to other usages of BTs. A merit of our variant is, however, that it has
deep analogies in the quantum case, as we shall see further.

\subsection{Duality}
\label{ss:duality}

Besides the $2\times2$ Lax matrix $L(u)$ which we used until now
there exists another,
$n\times n$ Lax matrix $\Lcal(v)$ for the Toda lattice
which is dual to $L(u)$ in the sense
that the corresponding spectral curves are equivalent up to interchanging
the spectral parameters $u$ and $v$
\beq
(-1)^{n-1}\det\bigl(u-\Lcal(v)\bigr)=\det(v-L(u)).
\label{eq:duality-L}
\eeq

Referring the reader to the paper \cite{AHH90} where the geometric meaning of
the duality is elucidated, we present here a more elementary approach.

To produce the dual Lax matrix $\Lcal(v)$ we take an eigenvector $\theta_1(u)$
of $L(u)$ corresponding to the eigenvalue $v$ (for brevity, we will
not mark the dependence on $u$ in $\theta$)
\beq
 L(u)\theta_1=v\theta_1
\label{eq:def-theta-1}
\eeq
and define by induction $\theta_j$ as
\beq
 \theta_{j+1}=\ell_j(u)\theta_j, \qquad j=1,\ldots,n.
\label{eq:def-theta-i}
\eeq

{}From \Ref{eq:def-theta-1} it follows that $\theta_{n+1}=v\theta_1$. The
function $\theta_j(u)$, when properly normalized,
is called {\it Baker-Akhiezer  function}.
Denoting the components  of the vector $\theta_j$ as $\varphi_j$ and $\psi_j$
we write down \Ref{eq:def-theta-i} explicitly as
\be
\left(\begin{array}{cc} \varphi_{j+1} \\ \psi_{j+1} \end{array}\right) =
\left(\begin{array}{cc} u+X_j & -\e^{x_j} \\ \e^{-x_j} & 0 \end{array}\right)
 \left(\begin{array}{cc} \varphi_j \\ \psi_j \end{array}\right).
\ee

Then, splitting the components and taking into account the quasiperiodicity
condition $\theta_{n+1}=v\theta_1$ we arrive at the following linear
equations for $\varphi_j$ and $\psi_j$:
\begin{subeqnarray}
 u\varphi_j&=&\varphi_{j+1}-X_j\varphi_j+\e^{x_j}\psi_j, \qquad
 j=1,\ldots,n-1 \\
  u\varphi_n&=&v\varphi_{1}-X_n\varphi_n+\e^{x_n}\psi_n,
\label{eq:eqs-phi}
\end{subeqnarray}
\begin{subeqnarray}
  \psi_{j+1}&=&\e^{-x_j}\varphi_j, \qquad j=1,\ldots,n-1 \\
 v\psi_1&=&\e^{-x_n}\varphi_n.
\label{eq:eqs-psi12}
\end{subeqnarray}

Eliminating $\psi_j$ we obtain a second-order finite-difference
equation for $\varphi_j$
\begin{subeqnarray}
  u\varphi_1&=&\varphi_2-X_1\varphi_1+\e^{x_1-x_n}v^{-1}\varphi_n,\\
  u\varphi_j&=&\varphi_{j+1}-X_j\varphi_j+\e^{x_j-x_{j-1}}\varphi_{j-1},
  \qquad j=2,\ldots,n-1 \\
  u\varphi_n&=&v\varphi_1-X_n\varphi_n+\e^{x_n-x_{n-1}}\varphi_{n-1},
\end{subeqnarray}
which can be rewritten as the linear problem for the
vector $\Phi$  with the components $\varphi_j$ in the matrix form:
\be
\Lcal(v)\Phi=u\Phi, \qquad
 \Phi=\left(\begin{array}{c} \varphi_1
\\ \ldots \\ \varphi_n \end{array}\right)
\ee
where the matrix $\Lcal(v)$ defined as
\beq
\Lcal(v)=\bmat{ccccc}
  -X_1 & 1 & \ldots & 0 & v^{-1}\e^{x_{1n}} \\
  \e^{x_{21}} & -X_2 & \ldots & 0 & 0 \\
  \ldots & \ldots & \ldots & \ldots & \ldots \\
  0 & 0 & \ldots & -X_{n-1} & 1 \\
  v & 0 & \ldots & \e^{x_{n,n-1}} & -X_n
     \emat, \qquad x_{jk}\equiv x_j-x_k
\label{eq:def-LL}
\eeq
is the dual Lax matrix we were looking for.

We leave the proof of the identity \Ref{eq:duality-L} as an exercise to the
reader. For the $r$-matrix corresponding to the Lax matrix $\Lcal(v)$ see
\cite{KSS98,Jim85}.

Similarly to the case of $2\times2$ matrix $L(u)$, for $\Lcal(v)$ there must
also exist a Darboux matrix $\Mcal$ intertwining $\Lcal(v;X,x)$ and
$\Lcal(v;Y,y)$. The explicit expression for $\Mcal$, like the one for
$\Lcal(v)$, can be found from the Baker-Akhiezer function. Let $\theta_j$
and $\tilde\theta_j$ refer, respectively, to $\Lcal(v;X,x)$ and $\Lcal(v;Y,y)$.
Let us assume that $\theta_j$ and $\tilde\theta_j$
are linked by the relation $\theta_j=M_j\tilde\theta_j$, which is obviously 
compatible with \Ref{eq:def-theta-i} and \Ref{eq:gauge-toda}.
Expanding $\theta_j=M_j\tilde\theta_j$ as
\be
 \bmat{c} \varphi_j \\ \psi_j \emat=
 \bmat{cc}
   u-\l+\e^{y_j-x_{j-1}} & -\e^{y_j} \\
   \e^{-x_{j-1}} & -1
 \emat
  \bmat{c} \tilde\varphi_j \\ \tilde\psi_j \emat,
\ee
taking its first line
\be
  \varphi_j=(u-\l+\e^{y_j-x_{j-1}})\tilde\varphi_j-\e^{y_j}\tilde\psi_j
\ee
and substituting
$u\tilde\varphi_j=\tilde\varphi_{j+1}-Y_j\tilde\varphi_j
                 +\e^{y_j-y_{j-1}}\tilde\varphi_{j-1}$
$  \tilde\psi_j=\e^{-y_{j-1}}\tilde\varphi_{j-1}$
from $\tilde\theta_{j+1}=\ell_j(u;Y_j,y_j)\tilde\theta_j$, as well as
$  Y_j=\e^{x_j-y_j}+\e^{y_j-x_{j-1}}-\l$
from (\ref{eq:XY-toda}b), we obtain, after making the necessary correction 
for $j=n$ the following result:
\begin{subeqnarray}
 \varphi_j&=&\tilde\varphi_{j+1}-\e^{x_j-y_j}\tilde\varphi_j,
 \qquad j=1,\ldots,n-1 \\
 \varphi_n&=&v\tilde\varphi_1-\e^{x_n-y_n}\tilde\varphi_n
\end{subeqnarray}
or, in matrix form, $\Theta=\Mcal\tilde\Theta$, with
\beq
  \Mcal(v)=\bmat{ccccc}
      -\e^{x_1-y_1} & 1 & \ldots & 0 & 0 \\
      0 & -\e^{x_2-y_2} & \ldots & 0 & 0 \\
      \ldots & \ldots & \ldots & \ldots & \ldots \\
      0 & 0 & \ldots & -\e^{x_{n-1}-y_{n-1}} & 1 \\
      v & 0 & \ldots & 0 & \e^{x_n-y_n}
    \emat.
\eeq

By construction, we have
\beq
  \Mcal(v)\Lcal(v;Y,y)=\Lcal(v;X,x)\Mcal(v).
\label{eq:McalL1}
\eeq

Alternatively, one could introduce $\tilde M_j\sim -M_j^{-1}$
\be
   \tilde M_j(u,\l)=\bmat{cc}
     1 & -\e^{y_j} \\
     \e^{x_{j-1}} & \l-u-\e^{y_j-x_{j-1}}
        \emat
\ee
such that
\be
  \tilde M_{j+1}(u,\l)\ell_j(u;X_j,x_j)
  =\ell_j(u;Y_j,y_j)\tilde M_j(u,\l)
\ee
and repeat the same calculation, starting from
$\tilde\theta_j=\tilde M_j\theta_j$.
The result is
$\tilde\Theta=\tilde\Mcal\Theta$,
with
\beq
  \tilde\Mcal(v)=\bmat{ccccc}
    1 & 0 & \ldots & 0 & -v^{-1}\e^{y_1-x_n} \\
    -\e^{y_2-x_1} & 1 & \ldots & 0 & 0 \\
      \ldots & \ldots & \ldots & \ldots & \ldots \\
    0 & 0 & \ldots & 1 & 0 \\
    0 & 0 & \ldots & -\e^{y_n-x_{n-1}} & 1
  \emat
\eeq
satisfying
\beq
  \tilde\Mcal(v)\Lcal(v;X,x)=\Lcal(v;Y,y)\tilde\Mcal(v)
\label{eq:McalL2}
\eeq

Despite the fact that $\tilde\Mcal\neq\Mcal^{-1}$ the formulas
\Ref{eq:McalL1} and \Ref{eq:McalL2} are compatible because of the
the remarkable factorization of $\Lcal(v)$:
\beq
  \Lcal(v;X,x)-\l\id=\Mcal(v)\tilde\Mcal(v), \qquad
  \Lcal(v;Y,y)-\l\id=\tilde\Mcal(v)\Mcal(v),
\eeq
see \cite{AM97,Ves91} for discussion of the factorization as a mechanism
for generating \bt{s}.

In the above formulas $v$ is, by definition, an eigenvalue of $L(u)$, so
the pair $(v,u)$ lies on the spectral curve $\det(v-L(u))=0$ of $L(u)$.
When dealing with $L(u)$, it is convenient to take $u$ as independent
variable, and when dealing with $\Lcal(v)$ respectively $v$. For \bt\ it means
in fact swapping the roles of $\l$ and $\mu$: the parameter $\mu$ becomes
independent numeric variable instead of $\l$. All the formulas defining
BT remain the same but their interpretation changes:
the equality \Ref{eq:def-mu-toda} becomes a constraint for $x$ and $y$
rather than definition of $\mu$, whereas $\l$ becomes a dynamical variable
--- a Lagrange multiplier for the constraint which can be determined from
equations \Ref{eq:XY-toda}. The respective dual \bt\ $\tilde\Bcal_\mu$
possesses all characteristic properties of BT which can be proven
using the Lax matrix $\Lcal(v)$ in the same manner as for  $\Bcal_\l$,
see \cite{KS5} for details.

\subsection{General construction of \bt}
\label{ss:reduction}

As shown in section \ref{ss:bt}, to any \bt\ $\Bcal_\l$ there corresponds
a Darboux matrix $M(u,\l)$ intertwining the corresponding Lax matrices,
see formula \Ref{eq:ML}. In practice, however, one usually does not know
the BT apriori, and has to deal with the inverse problem: given $L(u)$
how to find admissible $M(u,\l)$ producing a BT. If one is not interested
in the Hamiltonian properties of the transformation the usual strategy is
to try some ansatz for $M(u,\l)$, say, as a low-degree polynomial in $u$.
See the monograph \cite{MS91} for a plentitude of examples.

In this section we shall restrict our attention to the integrable models
generated by the quadratic Poissonian algebra \Ref{eq:Pb-LL} with the
$SL(2)$-invariant $r$-matrix \Ref{eq:r-toda} and address the following
question: which $M(u,\l)$ are admissible that is produce
canonical mappping $\Bcal_\l$?

{\bf Answer}: It is sufficient that $M(u,\l)$
\underline{as a smooth manifold} coincide with a symplectic
leave of the same quadratic Poisson bracket \Ref{eq:Pb-LL} as $L(u)$,
the leading coefficients $M^{(m)}$ of $M(u,\l)$ and
$L^{(n)}$ of $L(u)$ in $u$ commute:
\beq
[M^{(m)},L^{(n)}]=0
\label{eq:comML}
\eeq
and, also $M^{(m)}L^{(n)}\neq0$ (nondegeneracy condition).

{\bf Open problem:} Are these conditions necessary?

The \bt\ $\Bcal_\l$ constructed for the Toda lattice in section \ref{ss:bt}
also fits our scheme. Indeed, the Darboux matrix $M(u,\l)$ given by
\Ref{eq:def-M-toda} has, as a smooth manifold, the same structure as
the local Lax operator \Ref{eq:lDST} for the DST model. The parameter $\l$
is introduced through the shift $u\mapsto u-\l$ which is an automorphism of
the Poisson algebra.
To elaborate, let $M(u,\l)$ be
\beq
  M(u,\l;S,s)=\ell^{\tbox{DST}}(u-\l;S,s)
        \equiv\bmat{cc}
        u-\l+sS & -s \\ S & -1
          \emat.
\label{eq:MsS}
\eeq

As we shall see, the equation \Ref{eq:ML} allows then to determine
$S$, $s$ and, eventually, $L(u;Y,y)$ in terms of $L(u;X,x)$.
Expand first \Ref{eq:ML} in powers of $u$ using \Ref{eq:poly-L}
and \Ref{eq:MsS}.
The coefficient at $u^{n+1}$ vanishes because of \Ref{eq:comML}.
The matrix element $21$ of the coefficient at $u^n$
gives the expression for $S$:
\beq
   S=L_{21}^{(n-1)}(X,x).
\eeq

To determine $s$ take again \Ref{eq:ML} and substitute $u=\l$.
Multiplying the resulting matrix equality by the row-vector
$(1,-s)$ and noting that at $u=\l$ matrix $M(u,\l)$ degenerates:
\beq
  M(\l,\l)=\bmat{c} s \\ 1 \emat
       \begin{array}{cc}
         (S & -1) \\ {} & {} \end{array},
\eeq
we obtain the quadratic equation for $s$:
\beq
  L_{12}(\l;X,x)+s\bigl(L_{11}(\l;X,x)-L_{22}(\l;X,x)\bigr)
  -s^2L_{12}(\l;X,x)=0.
\eeq

Expressing the variables $S$ and $s$ in terms of $X$ and $x$ one can,
in principle calculate the Poisson brackets for $L(u;Y,y)$ directly and
verify that they have the same $r$-matrix form \Ref{eq:Pb-LL} as for
$L(u;X,x)$ proving thus the canonicity of the transformation from
$(X,x)$ to $(Y,y)$. See \cite{Skl51} where it is done in a slightly more
general situation. The calculation by brute force, however, is not
particularly instructive, and below, following \cite{Skl52},
we present a quite simple and general proof. The construction we describe 
is mimics the construction of the quantum $Q$-operator described in section
(\ref{ss:Q-toda}).

Suppose that $M(u)$ is a symplectic leaf of the same Poisson algebra
\Ref{eq:Pb-LL} as $L(u)$:
\beq
  \{\one M(u),\two M(v)\}=[r_{12}(u-v),\one M(u)\two M(v)]
\label{eq:pb-MM}
\eeq
satisfying the condition \Ref{eq:comML}.
Note that $M(u)$ is by no means
restricted to $\ell^{\tbox{DST}}(u-\l;S,s)$ as above.
Let $M(u)$ be parametrized by the canonical variables
$(S,s)$, and $L(u)$,  respectively, by $(X,x)$.
The matrix $M(u)$ might contain one or more parameters $\l$ which we neglect.
Assuming the commutativity \Ref{eq:comML}, consider two products:
$M(u)L(u)$ and $L(u)M(u)$. By virtue of the comultiplication
property of the bracket \Ref{eq:Pb-LL} they both are symplectic leaves
of the same bracket. Furthermore, due to the condition
$M^{(m)}L^{(n)}=L^{(n)}M^{(m)}\neq0$,
they share the same values of casimirs
described in section \ref{ss:qpb}, namely, the leading coefficient and
determinant. In a generic situation, provided there is no accidental
degeneration, which we shall assume, the equality of casimirs
implies an isomorphism of the symplectic leaves,
or, in other words, there should exist a canonical transformation
$\Lfrak:(X,x;S,s)\rightarrow(Y,y,T,t)$, determined from the equation
\beq
  M(u;T,t)L(u;Y,y)=L(u;X,x)M(u;S,s).
\label{eq:sx=yt}
\eeq

Suppose that the canonical transformation $\Lfrak$ has
a generating function $F(t,y;s,x)$
\beq
 X=\frac{\dd F}{\dd x}, \quad
 Y=-\frac{\dd F}{\dd y}, \quad
 S=\frac{\dd F}{\dd s},\quad
 T=-\frac{\dd F}{\dd t}
\label{eq:XYST}
\eeq
(for simplicity, we omit the indices $j$ in $X_j$, $x_j$ etc.)

Let us impose now the constraint
\beq
  t=s, \qquad T=S
\label{eq:constraint}
\eeq
and note that on the constraint surface we have $M(u;T,t)=M(u;S,s)$,
and therefore the equality \Ref{eq:sx=yt} is transformed to the Darboux
form \Ref{eq:ML}. It remains to prove that the transformation $\Lfrak$
remains canonical after being restricted on the constraint surface.

Suppose that one can resolve the equations
$\dd F/\dd s+\dd F/\dd t=0$
with respect to $s\equiv t$ and express $X$ and $Y$
from \Ref{eq:XYST}
as functions of $(x,y)$.

{\bf Proposition}. The resulting transformation
$\Bcal:(X,x)\rightarrow(Y,y)$
is canonical and is given by the generating function
$\Phi(x,y)=F\bigl(s(x,y),y;s(x,y),x\bigr)$, such that
\beq
  X=\frac{\dd \Phi}{\dd x}, \quad
 Y=-\frac{\dd \Phi}{\dd y}.
\eeq

{\bf Proof.} Let  $|_{st}$ mean the restriction on the constraint
manifold $s=t=s(x,y)$. The proof consists of two lines:
\bea
  X&=&\frac{\dd\Phi}{\dd x}=
  \left.\frac{\dd F}{\dd x}\right|_{st}
  +\left.\frac{\dd F}{\dd s}\right|_{st}\frac{\dd s}{\dd x}
  +\left.\frac{\dd F}{\dd t}\right|_{st}\frac{\dd t}{\dd x} \nonumber\\
  &=&\left.\frac{\dd F}{\dd x}\right|_{st}
  +\frac{\dd s}{\dd x}
  \left.\left(\frac{\dd F}{\dd s}+\frac{\dd F}{\dd t}\right)\right|_{st}.
\eea

We observe now that
\beq
  \left.\left(\frac{\dd F}{\dd s}+\frac{\dd F}{\dd t}\right)\right|_{st}
=0
\eeq
due to $S=T$, and, consequently, $X=\dd\Phi/\dd x$. Similarly, one
establishes  $Y=-\dd\Phi/\dd y$ completing thus the proof.

In many applications the Lax matrix $L(u)$, like in Toda case,
is a monodromy matrix factorized into the product of local Lax matrices
$\ell_j(u)$, see formula \Ref{eq:toda-Lll},
having the same Poisson brackets \Ref{eq:Pb-LL} as $L(u)$
\beq
  \{\one \ell_i(u),\two \ell_j(v)\}=
[r_{12}(u-v),\one \ell_i(u)\two \ell_j(v)]\d_{ij}.
\label{eq:pb-ll}
\eeq

The similarity transformation \Ref{eq:ML} is replaced now with
a gauge transformation \Ref{eq:gauge-toda}
which ensures the preservation of the spectral invariants of $L(u)$.

The modification of the reduction procedure described
above is quite straightforward. Supposing that $\ell_j(u)$ and
$M_j(u)$ depend on local canonical variables
we define first the local canonical transformations
$\Lfrak^{(i)}:(X_j,x_j;S_j,s_j)\rightarrow(Y_j,y_j,\allowbreak T_j,t_j)$
from the equations
\beq
  M_j(u;T_j,t_j)\ell_j(u;Y_j,y_j)
  =\ell_j(u;X_j,x_j)M_j(u;S_j,s_j).
\label{eq:Ml=lM}
\eeq

Let the corresponding generating functions be $f^{(j)}(t_j,y_j;s_j,x_j)$.
Consider the direct product of  $n$ phase spaces
$(X_j,x_j;S_j,s_j)$ and, respectively, $(Y_j,y_j;\allowbreak T_j,t_j)$.
The generating function
\beq
  F:=\sum_{j=1}^n f^{(j)}(t_j,y_j;s_j,x_j)
\eeq
determines then the direct product  $\Lfrak$ of the local
canonical transformations $\Lfrak^{(j)}$.

Let us now impose the constraint
\beq
  t_j=s_{j+1}, \qquad T_j=S_{j+1}
\label{eq:constr-chain}
\eeq
assuming periodicity  $j+n\equiv j$. The proof of the canonicity of the
resulting transformation ${\cal B}:(X,x)\rightarrow(Y,y)$
parallels the proof given previously. It remains to
notice that after imposing the constraint \Ref{eq:constr-chain}
we have $M_j(u;T_j,t_j)=M_j(u;S_j,s_j)$ and obtain the equality
\Ref{eq:gauge-toda}.

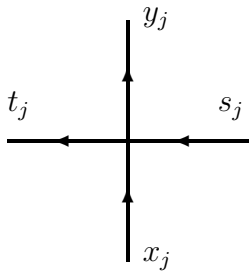
\begin{figure}[h]
\begin{picture}(30,9)(-15,0)
\thicklines
\put(4,0){\line(0,1){8}}
\put(0,4){\line(1,0){8}}
\put(4,2){\vector(0,1){.5}}
\put(4,6){\vector(0,1){.5}}
\put(2,4){\vector(-1,0){.5}}
\put(6,4){\vector(-1,0){.5}}
\put(4.5,0){$x_j$}
\put(4.5,8){$y_j$}
\put(0,5){$t_j$}
\put(7,5){$s_j$}
\end{picture}
\caption[tri]{Local transformation}
\label{fig:vertex}
\end{figure}

It is convenient to represent the structure of the \bt\ graphically.
Let the local transformation $\Lfrak^{(j)}$ be depicted as a
four-legged vertex (see figure \ref{fig:vertex}), each leg corresponding
to a canonical pair like $(X,x)$ etc. The arrows show the direction of
the transformation.

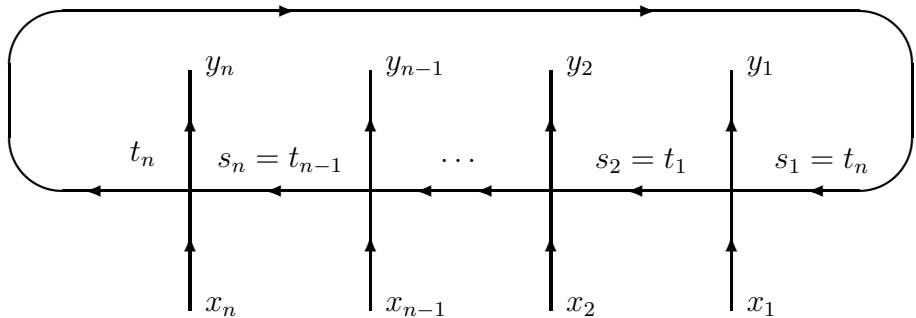
\begin{figure}[h]
\begin{picture}(30,10)(-4,-4)
\thicklines
\put(15,3){\oval(30,6)}
\put(6,-4){\line(0,1){8}}
\put(12,-4){\line(0,1){8}}
\put(18,-4){\line(0,1){8}}
\put(24,-4){\line(0,1){8}}
\put(6,0){\vector(-1,0){3.5}}
\put(12,0){\vector(-1,0){3.5}}
\put(18,0){\vector(-1,0){2.5}}
\put(18,0){\vector(-1,0){4.5}}
\put(24,0){\vector(-1,0){3.5}}
\put(27,0){\vector(-1,0){0.5}}
\put(6,6){\vector(1,0){3.5}}
\put(18,6){\vector(1,0){3.5}}
\put(6,-2){\vector(0,1){.5}}
\put(12,-2){\vector(0,1){.5}}
\put(18,-2){\vector(0,1){.5}}
\put(24,-2){\vector(0,1){.5}}
\put(6,2){\vector(0,1){.5}}
\put(12,2){\vector(0,1){.5}}
\put(18,2){\vector(0,1){.5}}
\put(24,2){\vector(0,1){.5}}
\put(6.5,-4){$x_n$}
\put(6.5,4){$y_n$}
\put(12.5,-4){$x_{n-1}$}
\put(12.5,4){$y_{n-1}$}
\put(18.5,-4){$x_2$}
\put(18.5,4){$y_2$}
\put(24.5,-4){$x_1$}
\put(24.5,4){$y_1$}
\put(4,1){$t_n$}
\put(9,1){\makebox(0,0){$s_n=t_{n-1}$}}
\put(15,1){\makebox(0,0){\ldots}}
\put(21,1){\makebox(0,0){$s_2=t_1$}}
\put(27,1){\makebox(0,0){$s_1=t_n$}}
\end{picture}
\caption[tri]{Composition of local transformations}
\label{fig:transfer-matrix}
\end{figure}

The \bt\ $\Bcal$ is represented then by the figure
\ref{fig:transfer-matrix} where the joint horizontal lines
mark the constraints \Ref{eq:constr-chain}.

In conclusion to this section, a few general remarks.
The proof of canonicity presented above
is pretty general using only the comultiplication property of the
quadratic $r$-matrix Poisson bracket \Ref{eq:Pb-LL}. It covers thus all
integrable models governed by the bracket
\Ref{eq:Pb-LL} for any $r$-matrix, not necessarily the $SL(2)$-invariant one.
The only thing one needs is to study
the structure of symplectic leaves of the Poisson
bracket and to choose some elementary matrices $M(u)$.

Note that the product of two $M$-matrices produces the composition of
corresponding \bt{s}. Given the conjecture about the factorization of
symplectic leaves from section \ref{ss:qpb} is true, it implies that
any BT is decomposable into elementary BTs corresponding to above
the Lax matrices
$\ell^{\tbox{XXX}}(u-\l)$,
$\ell^{\tbox{DST}}(u-\l)$,
$\check\ell^{\tbox{DST}}(u-\l)$, $\ell^{\tbox{Toda}}(u-\l)$,
$\check\ell^{\tbox{Toda}}(u-\l)$.

An interesting and yet unsolved problem is how to deal with the
spectrality property of BT within our construction. Our conjecture is that
there is a spectrality identity $\det(\e^\mu-L(\l))=0$
with respect to any zero $u=\l$ of $\det M(u)$.

\subsection{Application to Toda lattice}
\label{ss:bt-toda}

Let us demonstrate how the construction described in the previous section
produces the \bt\ for the Toda lattice described in section
\ref{ss:bt}. Substituting into the formula \Ref{eq:Ml=lM}
the expressions \Ref{eq:def-ell} for $\ell_j(u)$ and
\Ref{eq:MsS} for $M_j(u,\l)$ we get
\beq
  \bmat{cc} u-\l+t_jT_j & -t_j \\ T_j & -1 \emat
  \bmat{cc} u+Y_j & -\e^{y_j} \\ \e^{-y_j} & 0 \emat
=
  \bmat{cc} u+X_j & -\e^{x_j} \\ \e^{-x_j} & 0 \emat
  \bmat{cc} u-\l+s_jS_j & -s_j \\ S_j & -1 \emat
\eeq

The system of equations obtained by equating the coefficients at powers of
$u$ has a unique solution:
\begin{subeqnarray}
  Y_j&=&-\l+\e^{x_j} s_j^{-1}+s_jS_j, \\
  \e^{y_j}&=&s_j, \\
  T_j&=&\e^{-x_j}, \\
  t_j&=&\l\e^{x_j}-\e^{2x_j}s_j^{-1}+\e^{x_j} X_j.
\label{eq:solnYyTt}
\end{subeqnarray}
defining the local transformation $\Lfrak_\l^{(j)}$.
Strictly speaking, due to the degeneracy of the Lax matrix
$\ell^{\tbox{Toda}}(u)$
the proof of the canonicity of $\Lfrak_\l^{(j)}$ given in
section (\ref{ss:reduction})
does not apply here directly because the transformation
\Ref{eq:solnYyTt} does not possess a generating function in terms of
$(t,y,x,s)$. It is easy, however, to verify the canonicity by a straitforward
calculation.

The equalities (\ref{eq:solnYyTt}b) and (\ref{eq:solnYyTt}c) allow
to resolve the constraint \Ref{eq:constr-chain} yielding
$s_j=\e^{y_j}$, $S_j=\e^{-x_{j-1}}$ which, upon being substituted into
(\ref{eq:solnYyTt}a) and (\ref{eq:solnYyTt}d) produce exactly the defining
relations \Ref{eq:XY-toda} for the \bt\ studied in the section
\ref{ss:bt}.

{\bf Exercise.} Find what canonical transformations preserving the
Hamiltonians of the Toda lattice are generated by the Darboux matrices
$M(u)=\mbox{\rm diag}(1,a)$ and $M(u)=\ell^{\tbox{Toda}}(u-\l)$.

\section{Quantization}

\subsection{Quantum/classical correspondence}
\label{ss:quant}

Here we give only a very brief account of
the quantum mechanical notions we are going to use. For more information
on the basics of quantum mechanics see any good textbook.
See also Reshetikhin's lectures in this volume for references
on deformation quantization. 

The quantum observables are usually introduced as self-adjoint operators in
a Hilbert space. In the limit of the classical Hamiltonian mechanics,
as the deformation parameter $\hbar$ (Planck constant)
goes to $0$, the observables commute, and the next order term in $\hbar$
produces the Poisson bracket of the corresponding classical observables:
\be
   [\cdot,\cdot]=-\i\hbar\{\cdot,\cdot\}+O(\hbar^2).
\ee

We shall work with the realization of the Hilbert space of quantum states
as a space $\Lcal_2(\Rbbd^n)$ of square integrable functions of canonical
coordinates $(x_1,\ldots,x_n)$. The corresponding canonical momenta are
quantized then as the differentiation operators $X_j=-\i\hbar\dd/\dd x_j$.
Generally speaking, any operator $Q$ in $L_2(\Rbbd^n)$ can be realized as
an integral operator
\beq
  Q:f(x)\mapsto \int dx_1\ldots dx_n\,
     \Qcal(y\mid x)f(x)
\eeq
with the kernel $\Qcal(y\mid x)$ which possibly is a generalized function
(distribution).

To the canonical transformations in the classical mechanics (automorphisms
of the Poisson algebra) in the quantum mechanics there correspond the
automorphisms of the associative operator algebra, that is similarity
transformations $A\mapsto QAQ^{-1}$ with  unitary operators $Q$.
The following beautiful formula \cite{Fock}
\beq
  Q(y\mid x)\sim \exp\bigl(\i\hbar^{-1}F(y\mid x)\bigr), 
\qquad \hbar\rightarrow0
\label{eq:Fock}
\eeq
gives the correspondence between the kernel $Q(y\mid x)$ of a unitary
transformation and the generating function of the
classical canonical transformation into which it turns in the classical
limit. The formula \Ref{eq:Fock} works for non-unitary transformations
as well.

In what follows we shall occasionally use non-self-adjoint and non-unitary
operators which corresponds in the classical case to working with complex
rather than real manifolds.

\subsection{Quantum Toda lattice}
\label{ss:quantum-toda}

Starting from now we shall drop $\hbar$ from our formulas assuming $\hbar=1$.
The commutative Hamiltonians of the periodic quantum Toda lattice are 
differential operators in $\Lcal_2(\Rbbd^n)$. They are
obtained from exactly the same formulas \Ref{eq:def-H}, \Ref{eq:toda-Lll},
\Ref{eq:def-ell}, \Ref{eq:def-Hj} as in the classical case where one should
substitute $X_j=-\i\dd_{x_j}$. The proof of their commutativity is based
on the algebraic framework called {\it The Quantum Inverse Scattering Method},
see \cite{KBI93,Skl16,Skl32} for a detailed exposition of the method.

Starting with the quantum local Lax matrix $\ell_j(u)$
\be
 \ell_j(u)=\left(\begin{array}{cc}
     u-\i\dd_{x_j} & -\e^{x_j} \\
     \e^{-x_j} & 0 \end{array}\right)
\ee
we observe that it satisfies a quadratic commutation relation
\beq
  R_{12}(u_1-u_2)\one\ell(u_1)\two\ell(u_2)
  =\two\ell(u_2)\one\ell(u_1)R_{12}(u_1-u_2)
\label{eq:Rll}
\eeq
with
\beq
  R_{12}(u)=u+\i\Pcal_{12}.
\label{eq:def-R}
\eeq

As its classical counterpart \Ref{eq:Pb-LL}, the relation \Ref{eq:Rll}
possesses the comultiplication property which implies that the monodromy
matrix $L(u)$ given by \Ref{eq:toda-Lll} satifies the same relation
\beq
  R_{12}(u_1-u_2)\one L(u_1)\two L(u_2)=\two L(u_2)\one L(u_1)R_{12}(u_1-u_2)
\label{eq:RLL}
\eeq
from which the commutativity of the Hamiltonians
\be
   [t(u_1),t(u_2)]=0
\ee
follows immediately (see \cite{KBI93,Skl16,Skl32} for explanations).

The associative algebra given by the generators $L(u)$ and quadratic
relations \Ref{eq:RLL} is called yangian ${\cal Y}[gl_2]$. Its
representations correspond in the classical limit to the symplectic leaves
of the quadratic Poisson bracket \Ref{eq:Pb-LL}. A convenient way of viewing
the equality \Ref{eq:RLL} is to treat it as a particular form of the
quantum Yang-Baxter equation
\beq
  R_{12}(u)R_{13}(u+v)R_{23}(v)=R_{23}(v)R_{13}(u+v)R_{23}(v)
\label{eq:YBE}
\eeq
which is considered as an operator equality in the tensor product
$V_1\otimes V_2\otimes V_3$ of three linear spaces $V_1$, $V_2$ and $V_3$.
Respectively, $R_{jk}(u)$ is an operator in the space
$V_j\otimes V_k$ naturally embedded in $V_1\otimes V_2\otimes V_3$.
For any pair $V_j$, $V_k$ of the yangian ${\cal Y}[gl_2]$ moduli
there exists $R_{jk}$ such that the Yang-Baxter equation \Ref{eq:YBE} holds
for any triplet $V_1$, $V_2$, $V_3$.

In particular, for $V_1=V_2=V_3=\Cbbd^2$ we have the YBE \Ref{eq:YBE}
for the $R$-matrix \Ref{eq:def-R}. The relation \Ref{eq:RLL} also can be
considered as a particular case of \Ref{eq:YBE} for
$V_1=V_2=\Cbbd^2$ (auxiliary spaces),
$V_3=\Lcal_2(\Rbbd^n)$ (quantum space),
$R_{13}=\one L$, $R_{23}=\two L$.

The commutativity of the quantum Hamiltonians being established, the next
problem is to find an effective way of determining their joint spectrum.
There are two known ways of approaching this problem: separation of variables
\cite{Skl16,Sm-toda,KL99} and $Q$-operator \cite{Gau83,PG92}.
Here we shall consider the latter approach.

\subsection{Properties of $Q$-operator}
\label{ss:Q-prop}

The original idea of Baxter \cite{Bax72,Bax82}
which enabled him to solve the XYZ spin chain
was to construct a one-parametric family of operators $Q_\l$ commuting
with the Hamiltonians of the model
\beq
    [Q_\l,t(u)]=0
\label{eq:comm-Qt}
\eeq
and hence sharing with $t(u)$ the common set of eigenvectors. Moreover,
$Q_\l$ must satisfy the {\it Baxter equation}
\beq
   Q_\l t(\l)=\Delta_+(\l)Q_{\l+\i}+\Delta_-(\l)Q_{\l-\i}
\label{eq:Baxter-eq}
\eeq
where $\Delta_\pm(\l)$ are scalar functions determined by the parameters of
model. Note that in the left-hand-side of \Ref{eq:Baxter-eq} the order
$Qt$ or $tQ$ is not important because of the commutativity \Ref{eq:comm-Qt}.
Applying the Baxter equation \Ref{eq:Baxter-eq} to a common eigenvector of
$Q_\l$ and $t(\l)$ one can replace the operators in \Ref{eq:Baxter-eq}
by their eigenvalues. The resulting finite-difference equation of second
order for the eigenvalues of $Q_\l$ considered in an appropriate
functional class allows then to determine the spectrum of $t(\l)$.

Baxter succeeded to construct a $Q$-operator for the XYZ spin chain as a
trace of a monodromy matrix
\beq
   Q_\l=\tr_V \Lbbd_n(\l)\ldots\Lbbd_1(\l)
\label{eq:Q-LL}
\eeq
constructed with a specially chosen auxiliary space $V$. 
Graphically, the structure of $Q_\l$ is represented by the samefigure
\ref{fig:transfer-matrix} which we used in the classical case.
The horizontal lines correspond the auxiliary space, the vertical ones ---
to the quantum space. Each vertex represents an operator $\Lbbd_j(\l)$.
The commutativity
\Ref{eq:comm-Qt} is garanteed then by the Yang-Baxter equation, and
the only problem is to choose such $V$ which would produce the Baxter
equation \Ref{eq:Baxter-eq}.
Later Gaudin and Pasquier \cite{PG92} have constructed a $Q$-operator for
the quantum periodic Toda lattice by giving an explicit expression for
its kernel $\Qcal_\l(y\mid x)$ as an integral operator. They have also
noticed that the classical limit of the similarity transformation
$Q_\l(\cdot)Q_\l^{-1}$ is exactly the \bt\ studied in the previous sections.

Below we reproduce the result by Gaudin and Pasquier. Our approach combines
their integral operators technique with Baxter's original idea of
constructing as a trace of a monodromy matrix \Ref{eq:Q-LL}.

Note that such properties of \bt\ as the invariance of Hamiltonians
\Ref{eq:inv-H-cl} and spectrality \Ref{eq:spectrality} are the classical
counterparts of such properties of $Q$-operator as, respectively,
commutativity \Ref{eq:comm-Qt} and the Baxter equation \Ref{eq:Baxter-eq}.
The former one being obvious, we comment only the latter one.
Observing that the shift operators $\l\mapsto\l\mp\i$ are expressed as
$\exp(\mp\i\dd_\l)=\exp(\pm\mu)$ where $\mu$ is the canonical momentum
conjugate to $\l$ we can rewrite \Ref{eq:Baxter-eq} in the form
$t(\l)=\D_+(\l)\e^{-\mu}+\D_-(\l)\e^\mu$ which gives \Ref{eq:spectrality}
in the classical limit (for the Toda lattice $\D_\pm\equiv1$).

\subsection{$Q$-operator for Toda lattice}
\label{ss:Q-toda}

We shall construct the $Q$-operator as the trace of the monodromy matrix
\Ref{eq:Q-LL} taking for the auxiliary space $V$ the space $\Cbbd[s]$
of polynomials in variable $s$. The corresponding representation of the
yangian ${\cal Y}[gl_2]$ is realized then as the Lax operator of the
quantum DST model
\beq
  M(u,\l)=\bmat{cc} u-\l-\i\dd_s & -s \\ -\i\dd_s & -1 \emat,
\eeq
compare with the formula \Ref{eq:MsS} for the classical case.

To prove the commutativity \Ref{eq:comm-Qt} it is sufficient to establish
the identity
\beq
  M(u,\l)\ell(u)\Lbbd_\l=\Lbbd_\l\ell(u)M(u,\l)
\label{eq:MlL}
\eeq
which can be considered as a variant of the YBE \Ref{eq:YBE} with the
following layout of spaces: $V_1=\Cbbd^2$, $V_2=\Cbbd[s]$,
$V_3=\Lcal_2(\Rbbd^n)$. We shall use \Ref{eq:MlL} as the equation for
determining $\Lbbd_\l$. Rewriting \Ref{eq:MlL} as the system of equations
for the kernel $\Lbbd_\l(t,y\mid s,x)$ of $\Lbbd_\l$
\bea
 \lefteqn{
 \bmat{cc} u-\l-\i t\dd_t & -t \\ -\i \dd_t & -1 \emat
 \bmat{cc} u-\i \dd_y & -\e^y \\ \e^{-y} & 0 \emat
 \Lbbd_\l(t,y\mid s,x)
 }\nonumber\\
 &&=
 \bmat{cc} u+\i \dd_x & -\e^x \\ \e^{-x} & 0 \emat
 \bmat{cc} u-\l+\i +\i s\dd_s & -s \\ \i \dd_s & -1 \emat
 \Lbbd_\l(t,y\mid s,x)
\eea
we obtain a unique, up to a scalar factor, solution
\beq
 \Lbbd_\l(t,y\mid s,x)\sim
 \d(s-\e^y)\exp\bigl(\i t\e^{-x}-\i \e^{x-y}+\i \l(x-y)\bigr).
\label{eq:Lbbd}
\eeq

From \Ref{eq:Q-LL} we get the formula for the kernel of $Q_\l$:
\beq
  \Qcal_\l(y\mid x)=\int ds_n\ldots\int ds_1 \prod_{j=1}^n
  \Lbbd_\l(s_{j+1},y_j\mid s_j,x_j).
\label{eq:ker-Q}
\eeq

The integration over $s_j$ in \Ref{eq:ker-Q} reduces, due to the
delta-function factor in \Ref{eq:Lbbd}, to the substitution
$s_j=\e^{y_j}$. Finally, we have
\beq
  \Qcal_\l(y\mid x)= \prod_{j=1}^n
  \exp\bigl(\i \e^{y_{j+1}-x_j}-\i \e^{x_j-y_j}+\i \l(x_j-y_j)\bigr).
\label{eq:kerQ}
\eeq

Note that $\Qcal_\l(y\mid x)=\exp\left(-\i F_\l(y\mid x)\right)$ where
$F_\l(y\mid x)$ is the generating function \Ref{eq:F-toda} of the classical BT,
that is the semiclassical formula \Ref{eq:Fock} is exact in our case. 
This is an accidental peculiarity of Toda lattice which usually
does not hold for other models.

In \cite{PG92} another version of the $Q$-operator is used which differs from
\Ref{eq:kerQ} by the shift $y_j\mapsto y_j+\i\pi/2$
\beq
  \check\Qcal_\l(y\mid x)= \prod_{j=1}^n
  \exp\bigl(-\e^{y_{j+1}-x_j}-\e^{x_j-y_j}+\l(x_j-y_j)\bigr)
\label{eq:kerQ1}
\eeq
which, in operator terms, corresponds to multiplying $Q_\l$ by the factor
$\exp(-\pi H_1/2)$. The kernel \Ref{eq:kerQ1} is more convenient for analytical
study since it rapidly decreases along the real axis in $x_j$.

\subsection{Baxter's equation}
\label{ss:Baxter-eq}

The commutativity \Ref{eq:comm-Qt} being already established, it remains to
prove for our $Q_\l$ the Baxter equation \Ref{eq:Baxter-eq}.
We reproduce here the proof by Gaudin and Pasquier \cite{PG92}
which parallels the proof for the classical case given in the end of
section \ref{ss:bt}, see formula \Ref{eq:spectrality}.

First, note that the kernel \Ref{eq:kerQ} factorizes as
\beq
  \Qcal_\l(y\mid x)=\prod_{j=1}^n w_j(\l)
\eeq
into factors
\beq
  w_j(\l)=\exp\bigl(\i \e^{y_j-x_{j-1}}-\i \e^{x_j-y_j}
          +\i \l(x_{j-1}-y_j)\bigr)
\eeq

Applying then $t(\l)$ to the kernel $\Qcal_\l(y\mid x)$ and
using \Ref{eq:toda-Lll} we observe that each $\ell_j(\l;-\i\dd_{y_j},y_j)$
acts locally only on $w_j(\l)$ and obtain
\beq
 t(\l)\Qcal_\l(y\mid x)=
    \tr\bigl(\ell_n(\l)w_n(\l)\bigr)\ldots\bigl(\ell_1(\l)w_1(\l)\bigr)
                   =Q_\l(y\mid x)\tr\tilde\ell_n\ldots\tilde\ell_1
\eeq
where
\beq
 \tilde\ell_j\equiv \ell_j(\l)\ln w_j(\l)
  =\bmat{cc}
    \e^{y_j-x_{j-1}}+\e^{x_j-y_j} & -\e^{y_j} \\
    \e^{-y_j} & 0
   \emat.
\eeq

After that we can use the triangular gauge transformation
$\widehat\ell_j\equiv N_{j+1}^{-1}\tilde\ell_jN_j$
with $N_j$ and the resulting matrix $\widehat\ell_j$
given by the same formulas \Ref{eq:triangN} as in the classical case.
Noticing then that
\beq
  \frac{w_j(\l+\i )}{w_j(\l)}=\e^{y_j-x_{j-1}}, \qquad
  \frac{w_j(\l-\i )}{w_j(\l)}=\e^{x_{j-1}-y_j}
\eeq
we obtain the required result
\be
  t(\l)Q_\l=Q_{\l+\i }+Q_{\l-\i }
\ee

Similarly, for the modified kernel \Ref{eq:kerQ1} one obtains
\beq
  t(\l)Q_\l=\i^nQ_{\l+\i }+\i^{-n}Q_{\l-\i }.
\label{eq:Baxter-eq-toda}
\eeq

The Toda Hamiltonians $\{H_j\}_{j=1}^n$ enter the Baxter equation 
\Ref{eq:Baxter-eq-toda} through the generating function
$t(u)=u^n+H_1u^{n-1}+\ldots+H_n$.
Their eigenvalues are determined by the condition that the finite-difference
equation
\Ref{eq:Baxter-eq-toda} possesses a solution $Q_\l$ which is holomorphic
and rapidly decreases along the real axis. For a detailed analysis of the
equation \Ref{eq:Baxter-eq-toda} see \cite{PG92,KL99}.

\section{Conclusion}

We have discussed here only the most elementary properties of
\bt\ and $Q$-operator using the sole example of Toda lattice.
For further reading see \cite{BLZ,Der,KL99,KS5,KSS98,KV00,Pro,Sm-toda,Sm-rt}.

I am grateful to the University of Montreal and CRM for the hospitality and
the opportunity to put together these lectures.
My deep thanks are addressed
 also to my coathor Vadim Kuznetsov the collaboration
with whom provided material for these lecturues.


\end{document}